# Publishing and Discovery of Mobile Web Services in Peer to Peer Networks

Satish Narayana Srirama

Department of Computer Science, Informatik V
RWTH Aachen University, Germany
srirama@cs.rwth-aachen.de

**Abstract:** It is now feasible to host Web Services on a mobile device due to the advances in cellular devices and mobile communication technologies. However, the reliability, usability and responsiveness of the Mobile Hosts depend on various factors including the characteristics of available network, computational resources, and better means of searching the services provided by them. P2P enhances the adoption of Mobile Host in commercial environments. Mobile Hosts in P2P can collaboratively share the resources of individual peers. P2P also enhances the service discovery of huge number of Web Services possible with Mobile Hosts. Advanced features like post filtering with weight of keywords and context-awareness can also be exploited to select the best possible mobile Web Service. This paper proposes the concept of Mobile Hosts in P2P networks and identifies the means of publishing and discovery of Web Services in mobile P2P networks.

## 1 Introduction

Traditionally, the hand-held devices have many resource limitations like low computation capacities, limited storage capacities, and small display screens with poor rendering quality. Most recently, the capabilities of these wireless devices like smart phones, PDAs are expanding quite fast and this is resulting in their quick adoption in domains like mobile banking, location based services, e-learning etc. The situation also brings out a large scope and demand for software applications for such high-end wireless devices.

Moreover, with the achieved high data transmission rates in cellular domain, with interim and third generation mobile communication technologies like GPRS, EDGE and UMTS [GSM], mobile phones are also being used as Web Service clients and providers, bridging the gap between the wireless networks and the stationery IP networks. Web Services have a broad range of service distributions and on the other hand cellular phones have large and swiftly expanding user base. Combining Web Services with mobile technology brings us a new trend and lead to manifold opportunities to mobile operators, wireless equipment vendors, third-party application developers, and end users. [JSR, SJP06a, Ba03]

While mobile Web Service clients are quite common these days, the research with mobile Web Service provisioning is still sparse. To support this, during one of our previous projects, a small mobile Web Service provider ("Mobile Host") has been developed for resource constrained smart phones. The detailed performance analysis conducted with the Mobile Host showed that the processing capability, time frames are very much within the acceptable levels. [SJP06a]

Subsequently, in recent years, peer-to-peer technology is being used in vast applications in domains like entertainment systems, ubiquitous computing, pervasive computing and collaborative systems etc. P2P is gaining popularity as low-cost individual computing technology. Three main classes of applications have emerged in the P2P environment over time. 1. Content and file management P2P applications like Napster [Nap] and Gnutella [Gnu]. 2. Parallelizable P2P applications that split large tasks into smaller chunks that execute in parallel over autonomous peers like SETI@Home [SETI]. 3. Collaborative P2P applications that allow users to collaborate with each other with out the help of central servers to collect and relay information, like Skype [Skype]. Most recently, ad-hoc networks of mobile terminals are also participating in such P2P networks and applications like in Magi [Bo00].

It is quite interesting, to combine these two trends and to come up with a scenario where mobile Web Services being used in P2P mobile network also, further increasing the application scope of both P2P and mobile Web Services. Not just the enhanced application scope, the merging can also provide better options for service discovery of huge number of Web Services possible by providing them from Mobile Hosts. Our current research focuses at realizing the Mobile Hosts in P2P networks and tries to find alternatives for service discovery using techniques like advanced service discovery with weights for keywords and context-awareness. The paper is organized as follows.

Section 2 discusses the concepts of mobile Web Service provisioning. Section 3 gives a brief description of mobile Web Services in P2P systems, addressing the convergence of Web Services and P2P systems. Section 4 discusses the details of mobile Web Service publishing and discovery in JXTA network. Section 5 refines the basic discovery mechanism with advanced features like weight of keywords and context-awareness. Section 6 concludes the paper and proposes future research directions.

## 2 Mobile Web Service provisioning

Service Oriented Architecture (SOA) [Bu00] is the latest trend in information systems engineering. It is a component model, presenting an approach for building distributed systems. SOA delivers application functionality as services to end-user applications and other services, bringing the benefits of loose coupling and encapsulation to the enterprise application integration. SOA defines participating roles as, service provider, service client, and service registry. Figure 1 shows the SOA collaborations. The operations publish, find, bind and the artifacts services and descriptions of SOA are also shown in the figure 1. SOA is not a new notion and many technologies like CORBA and DCOM are at least partly represent this idea. Web Services are newest of these developments and by far the best means of achieving SOA.

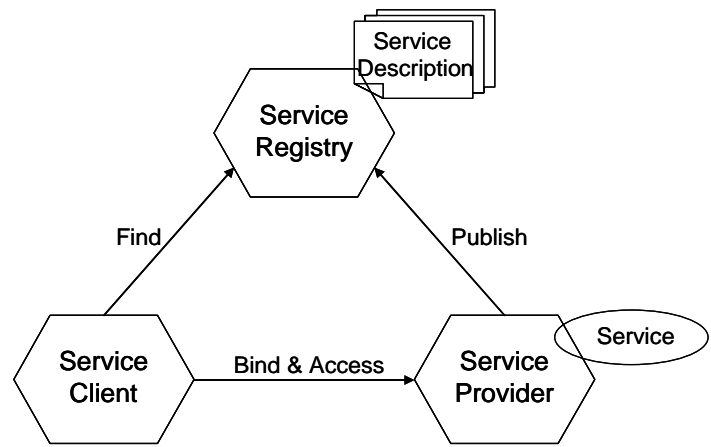

Figure 1: Service Oriented Architecture collaborations

The Web Service architecture defined by the W3C [W3Ca] enables application-to-application communication over the Internet. Web Services are self-contained, modular applications whose public interfaces are described using Web Services Description Language (WSDL) [W3Cb]. Web Services allow access to software components through standard Web technologies and protocols like SOAP [W3Cc] and HTTP [IETF99], regardless of their platforms, implementation details. A service provider develops and deploys the service and publishes its description and binding/access details (WSDL) with the UDDI [UDDI] registry. Any potential client queries the UDDI, gets the service description and accesses the service using SOAP. The communication between client and UDDI registry is also based on SOAP.

Web Services and its protocol stack are based on open standards and are widely accepted over the internet community. Web Services have wide range of applications and range from simple stock quotes to pervasive applications using context-awareness like weather forecasts, map services etc. The biggest advantage of Web Services lies in its simplicity in expression, communication and servicing. The componentized architecture of Web Services also makes them reusable, thereby reducing the development time and costs.

The quest for enabling these open XML Web Service interfaces and standardized protocols also on the radio link lead to new domain of applications mobile Web Services. In this domain, the resource constrained mobile devices are used as both Web Service clients and providers. During one of our previous projects, we have developed and analyzed the performance of a mobile Web Service provider for smart phones.

Mobile Host is a light weight Web Service provider built for resource constrained devices like cellular phones. It has been developed as a Web Service handler built on top of a normal Web server. The Web Service requests sent by HTTP tunneling are diverted and handled by the Web Service handler. The Mobile Host was developed in PersonalJava [JCP00] on a SonyEricsson P800 smart phone. The footprint of our fully functional prototype is only 130 KB. Open source kSOAP2 [Ksoap] was used for creating and handling the SOAP messages.

The detailed evaluation of this Mobile Host clearly showed that service delivery as well as service administration can be done with reasonable ergonomic quality by normal mobile phone users. As the most important result, it turns out that the total WS processing time at the Mobile Host is only a small fraction of the total request-response time (<10%) and rest all being transmission delay. This makes the performance of the Mobile Host directly proportional to achievable higher data transmission rates.

The second-generation GSM networks delivered high quality and secure mobile voice and data services like SMS (Short Message Service), circuit switched Internet access etc., with full roaming capabilities and across the world. The GSM platform is a widely successful wireless technology and it is the world's leading mobile standard. But, with the advent of the interim-generation technologies like GPRS [GPRS] and EDGE [EDGE], and third-generation technologies like UMTS [UMTS], still higher data transmission rates are achieved in the wireless domain, in the order of few hundreds of Kbs to 2 Mbs. Most recently with the advent of 4G technologies and their deployment in south Asian countries suggests that mobile data transmissions of the rate of few GB is also possible [4G05]. These developments make the Mobile Host soon realizable in commercial environments.

Mobile Host opens up a new set of applications and it finds its use in many domains like collaborative learning, social systems, mobile community support and etc. Many applications were developed and demonstrated using Mobile Host, for example in a distress call, the mobile terminal could provide a geographical description of its location along with location details. Another interesting application scenario involves the smooth co-ordination between journalists and their respective organizations. From a commercial viewpoint, Mobile Host also renders possibility for small mobile operators to set up their own mobile Web Service business without resorting to stationary office structures.

# 3 Mobile Web Services in P2P networks

Once the Mobile Host was developed, and its feasibility analyzed, extensive study was conducted in finding its specific application domains. The study was also aimed at growing Mobile Hosts' application scope; our research mainly focused on mobile community support and pervasive applications. During this study, it was observed that most of the targeted collaborative applications, somehow converged to P2P applications and P2P offered a large scope for many applications with Mobile Host. Not just the enhanced application scope, the P2P also offers many technical advantages to the Mobile Host. Hence our current research in this domain focuses at checking the feasibility of Mobile Host in the P2P world.

## 3.1 Convergence of P2P and Web Services

P2P is a set of distributed computing model systems and applications used to perform a critical function in a decentralized manner. Peers are autonomous. In its pure form; each peer acts as both server and client. P2P takes advantage of resources of individual peers like storage space, processing power, content and achieves scalability, cost sharing and anonymity, and thereby enabling ad-hoc communication and collaboration. P2P systems have evolved across time and have wide range of applications and provide a good platform for many data and compute intensive applications. [Nap, Gnu, SETI, Skype]

The first generation P2P systems like Napster used centralized servers for maintaining an index of the connected peers and their resources. The indexes can later be queried by the peers and the resources are downloaded from the providers using IP networks. But these centralized systems have single points of failure and produce giant communication traffic and storage on server resulting bottlenecks. These drawbacks lead to the second generation of P2P systems like Gnutella which used a complete decentralized network. Unlike Napster, Gnutella would connect users directly to a group of other users and so on. For this, Gnutella uses pre-existing, extendable list of possible working peers, whose addresses are embedded inside the application code. But these decentralized networks formed islands in the P2P network and their search functions were unreliable and may not query entire network. The third generation P2P systems like eDonkey [ED00] and Bit Torrent [BT06] are a hybrid of the previous two generation technologies and made enhancements to improve their ability to deal with large numbers of users using concepts like super peers. Super peers have higher resource capabilities and act as relays for other peers and super peers. Super peers also have abilities to traverse NAT and firewall.

P2P systems can also leverage SOA and are generally designed to enable loosely coupled systems. The concept of services and the similarities of description stack of both P2P and Web Services make them comparable [Sc01]. The major difference being; Web Services will be well-known hosts with static IP addresses, and are based on a centralized model and primarily focused on standardizing messaging formats and communication protocols. P2P systems, on the other hand, are based on a decentralized model and primarily focused on supplying processing power, content, or applications to peers in a distributed manner, and less focused on the semantics of messaging formats and communication protocols. In the P2P world the peers jump through potential jumble of firewalls, NATs and proxies trying to connect to other peers.

**3.2 Mobile Host in JXME network**

In order to adapt the Mobile Host to the P2P network, many of the current P2P technologies like Gnutella, Napster and Magi are studied in detail. Most of these technologies are proprietary and are generally targeting specific applications. Only Project JXTA [JXTA] offers a language agnostic and platform neutral system for P2P computing. JXTA technology is a set of open protocols that allow any connected device on the network ranging from cell phones and wireless PDAs to PCs and servers to communicate and collaborate in a P2P manner. JXTA enables these devices running on various platforms not only to share data with each other, but also to use functions of their respective peers. JXTA peers use XML as standard message format and create a virtual P2P network over these devices connected over different networks.

Moreover the JXTA community has developed a light version of JXTA for mobile devices, called JXME (JXTA for J2ME). JXME works on MIDP supporting devices like smart phones. JXME simplifies the Mobile Host's entry to P2P domain. JXME has two versions: proxyless and proxied. The proxyless version works similar to native JXTA, whereas the proxied version needs a native JXTA peer to be set up as its proxy. The proxied version is lighter of the two versions and peers using this version participate in binary communication with their proxies.

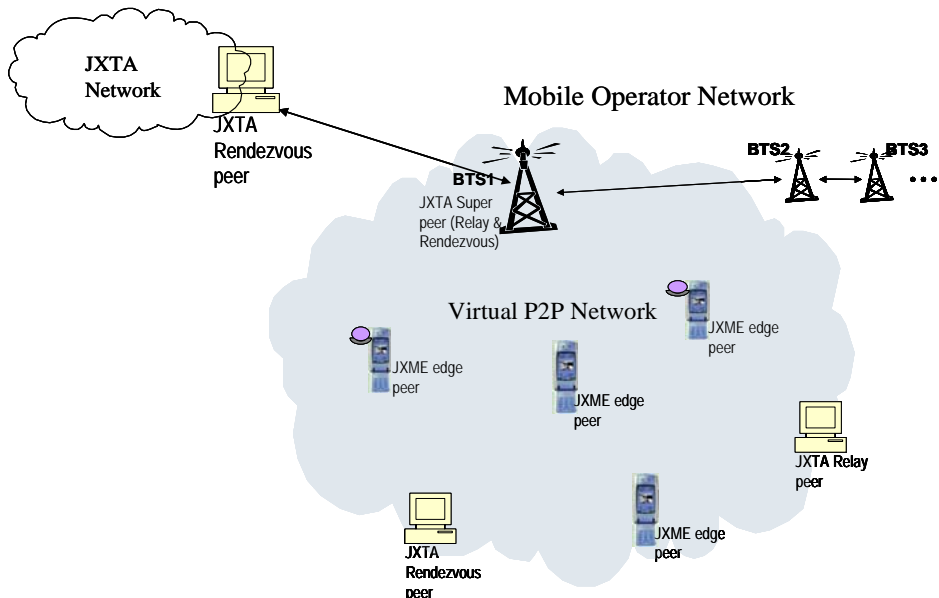

Figure 2: General Architecture of Mobile Hosts in JXTA network

JXME provides a perfect platform for Mobile Hosts entry to P2P networks. Considering JXTA also eliminates many of the low level details of the P2P systems like the transportation details. The peers can communicate with each other using the best of the many network interfaces supported by the devices like Ethernet, WiFi etc. Moreover JXTA dynamically uses either TCP or HTTP protocols to traverse network barriers, like NATs and firewalls.

Figure 2 shows the general architecture of Mobile Hosts in the JXME network. The virtual P2P network is established in the mobile operator network with one of the node in operator proprietary network, acting as a JXTA super peer. The super peer can exist at Base Transceiver Station (BTS) and can be connected to other base stations extending the JXTA network into the mobile operator network. Any Mobile Host or mobile Web Service client in the wireless network can connect to the P2P network using the node at base station as the rendezvous peer. The super peer can also relay requests to and from JXTA network, to the smart phones. With in this network, the participating smart phones can be addressed with both peer ID and the mobile phone number. Standalone systems can also participate in such a network as both rendezvous and relay peers, if the operator network allows such functionality, further extending the network.

Mobile Host in JXME network offers many advantages in domains like collaborative learning, image sharing, and location based services etc, taking advantage of individual peers' resources like storage space, processing power. Moreover, the mobile phone users in the operator network might not use the Web Services for the development purpose. General mobile users are interested in applications rather than individual components or Web Services. An application might use one or more Web Services at the backend and can be provided as an installable application. In such a situation, the P2P network can offer easy means of storing and sharing these installable client applications for the participating peers.

Not just the enhanced application scope, the JXME network also provides many technical advantages to the Mobile Host like enhanced service discovery and access mechanisms. With in JXTA network, each peer is uniquely identified by a static peer ID, which allows the peer to be addressed independent of its physical address like DHCP based IP address. This peer ID will stay forever with that device even though the device supports multiple network interfaces like Ethernet, WiFi for connecting to the P2P network. By using peer ID, Mobile Host does not have to worry about changing IPs, operator networks, and is always visible to the Web Service client. Mapping the peer ID to the IP is taken care by the JXTA network, thus eliminating the need for public IP. The public IP for each of the participating Mobile Hosts was observed to be the major hindrance for commercial success of Mobile Host. [SJP06b]

# 4 Publishing and discovery of mobile Web Services

Typically, Web Services are built for static networks and are published to a UDDI registry. Any potential client searches for the Web Service in the registry, gets the WSDL of the service and then accesses the Web Service using the information specified by the WSDL document. When we consider service discovery and registry for mobile Web Service provisioning, the numerous Web Services possible with each smart phone user providing some services in the mobile operator network, may make the centralized UDDI registry not the best solution.

Besides, mobile networks are dynamic due to node movement. Nodes can join or leave network at any time and can switch from one operator to another operator over the network. This might make the binding information in the WSDL documents, inappropriate. Keeping up to date information of the published services in central registries is really difficult. JXTA can provide alternatives for these problems. JXTA can scale by collaboratively using the resources of individual peers. Module advertisements provide alternatives for mobile Web Service discovery.

## 4.1 Publishing mobile Web Services in JXTA network

All resources like peers, peer groups and the services provided by peers in JXTA network are described using Advertisements. Advertisements are language-neutral metadata structures represented as XML documents. Peers discover resources by searching for their corresponding advertisements, and may cache any of the discovered advertisements locally. Every advertisement exists with a lifetime that specifies the availability of that resource. Lifetimes gives the opportunity to control out of date resources without need of any centralized control mechanism. To extend the life time of an advertisement, the resource can be republished.

Similarly, Web Services deployed on Mobile Host in the JXTA network are to be published as JXTA advertisements, so that they can be sensed as JXTA services among other peers. JXTA specifies 'Modules' as a generic abstraction that allows peers to describe and instantiate any type of implementation of a behavior in the JXTA world. So the mobile Web Services are published as JXTA modules in the P2P network. The module abstraction includes a module class, module specification, and module implementation. The module class is primarily used to advertise the existence of a behavior. Each Module class contains one or more module specifications, which contain all the information necessary to access or invoke the module. The module implementation is the implementation of a given specification. There might be more than one implementation for a given specification across different platforms.

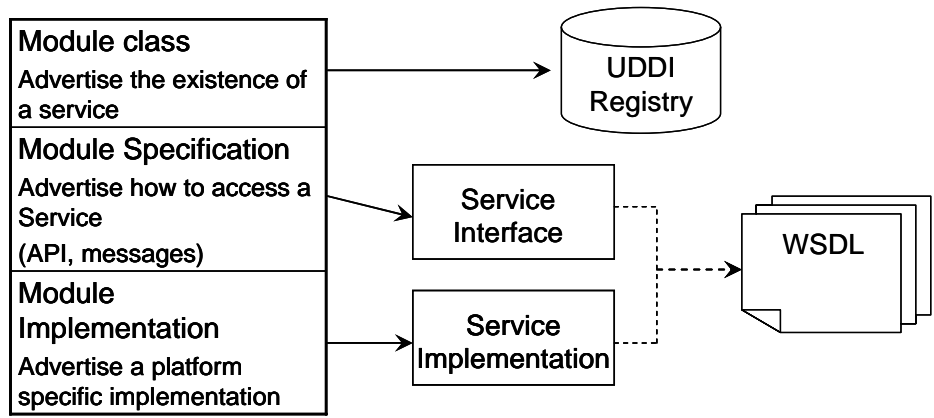

Figure 3: Mapping between JXTA Modules and Web Services

Figure 3 shows the mapping between JXTA Modules and Web Services. The collection of module abstractions represent the UDDI in a sense of publishing and finding service description and WSDL in a sense of defining transport binding to the service. Further information of WSDL can be added as extra parameters to module class advertisements. Any peer can query these module advertisements for the Web Services provided by the Mobile Hosts and can accesses these services across the JXTA networks. The Web Services provided by the Mobile Hosts, can also be classified using peer groups. The modules are organized in different peer groups according to categorizations. [NT05]

**4.2 Mobile Web Service discovery in JXTA network**

The JXTA Modules are searchable and the advertisements can be regularly re-published, always maintaining the current state of the Web Services. Lifetimes of advertisements, comes in handy and out of date resources can be deleted automatically without need of any centralized control mechanism. Thus Modules handle the dynamic nature of mobile Web Services.

Web Services represented as Modules at JXTA network can be searched by name and description parameters. The JXTA API provides a simple keyword search on the name and description of the Modules advertised in mobile P2P network. As we are considering about huge numbers of Web Services, these basic parameters might not be sufficient to find out the exact search results. In fact, some valuable information like context information may not be included in the name and descriptions, as the search is purely text based. Moreover we would like to extend the search criteria to the WSDL level. This means that search parameters would not be restricted to Module class advertisement details. The search will also extend by looking up the WSDL tags and information. Advanced discovery of mobile Web Services in P2P is also being addressed.

This detailed search mechanism might not be performed at the JXME edge peer because of the resource limitations of the smart phones like low computational capabilities and limited storage capabilities. So this advanced search mechanism can be shifted to a standalone distributed middleware. In this domain, we are trying to realize an Enterprise Service Bus (ESB) [Sc03] based "Mobile Web Services Mediation Framework" (MWSMF) [SJP06c], which maintains the individual user profiles, personalization settings and context sensitive information. ESBs are the emerging infrastructure components for realizing SOA and enterprise integration. In the scenario where the Mobile Host uses the proxied version of JXME, the proxy node can be a participant in the mediation framework.

Modules advertising the Web Services in JXTA can also be properly categorized using peer groups. Web Services of the same category like services of same publisher, same business can thus be published in the same peergroups. Categories help in identification or classification of all the Web Service types and help in easy discovery of Web Services. The peergroups thus simulate the tModel feature of the UDDI.

## 5 Advanced mobile Web Service discovery

The basic mobile Web Service discovery in JXTA networks, across Module advertisements is purely based on text based keywords. So the search might return a large number of resulted services based on keyword matching. In our current research, we are trying to refine this search with techniques like post-filtering with weight of keywords and context-awareness.

### 5.1 Post-filtering of mobile Web Services

Once the JXTA API returns the search results for Web Services, the post-filtering query feature can be applied to help find out most relevant advertisements among the currently available ones. The service advertisements (Modules) and descriptions (WSDL) obtained from the basic search are cached at the local repository. The advertisements can later be filtered with weight of the keywords. The main idea behind this approach is that people usually express their opinion by using frequently used words.

The filtering algorithm is based on the vector space model. The documents and query are represented on a K-dimensional vector space. K is the number of distinct words in the document collection. Each word is assigned a TF-IDF [SB88] weight (term frequency–inverse document frequency), which statistically reflects the importance of the word to the document. The value is calculated based on the word's frequency and its distribution across a collection of documents. The importance is directly proportionally to the number of times a word appears in the document and is inversely proportional to how common the word is in the collection of the documents. TF-IDF is often used by search engines to find the most relevant documents to a search query. The similarity of two documents can also be calculated based on TF-IDF and the cosine similarity between the angles of the two vectors which represent the documents. This value is then normalized 0 through 1, and is used to rank the search results. [Th05]

The Web Service search results obtained after performing the TF-IDF weight search represent a small subset of the total service advertisements and descriptions, cached at local repository after JXTA module search. The subset can be identified as Advanced Matching Services (AMS). But as the target of the search results is a resource constrained device like smart phone, with little displays and poor rendering capabilities, the search can further be refined with context-aware service discovery. The final target of this advanced search is to provide only the best matching services to the mobile. The mobile user can scroll through the list and can select the best possible Web Service. The user can then access the Web Service from the respective Mobile Host in the JXTA network, by downloading and installing the client software from the P2P network.

**5.2 Context-aware service discovery**

The context-aware service discovery provides the most appropriate and relevant services for the mobile Web Service clients. The context is the information that can be used to characterize the situation of an entity. An entity is a person, place, or object that is considered relevant to interaction between a user and an application including the user and application themselves. Context-awareness is a property of a system that uses context to provide relevant information and/or service to the user, where relevancy depends on the user's task. Thus, context-aware service discovery can be defined as the ability to make use of context information to discover the most relevant services for the user. [De00]

Mobile Web Service clients generally prefer using services of the Mobile Host based on several context parameters such as location, time, device capabilities, profiles, and load on the Mobile Host etc. The service context can be described using ontology-based mechanism. For describing the semantics of services, the latest research in service-oriented computing recommends the use of Web Ontology Language (OWL) [W3Cd] based Web Ontology Language for Services (OWL-S) [W3Ce]. OWL-S is an ongoing effort to enable automatic discovery, invocation, and composition of Web Services. The ontology is composed of ServiceProfiles describing the capabilities of the services like inputs and outputs, ServiceGroundings describing the invocation details of the services like communication scheme, address, ports, etc., and Service-Models describing the tasks and behavior of the services. The ServiceProfile describes the functional and non-functional aspects of a Web Service, and is therefore used for mobile Web Service discovery. The ontologies can later be processed during the service matching. [EB06]

The context source provides the context information of the services and Web Service clients. The Web Service description contains the reference to its context source. The context engine matches the suitable services for the mobile Web Service client by processing the context information of advanced matching services and the context of mobile Web Service client itself. The collected context information of the AMS will be stored in services graph structure. The context engine uses semantic match making algorithms to obtain the most suitable services. We are planning to use Jena [Jena] for semantic matchmaking. Jena includes a rule-based inference engine, support for ontologies, a querying mechanism and a persistent storage capability. The context source and the service matching can also be maintained at the MWSMF along with post-filtering features of the advanced mobile Web Services discovery.

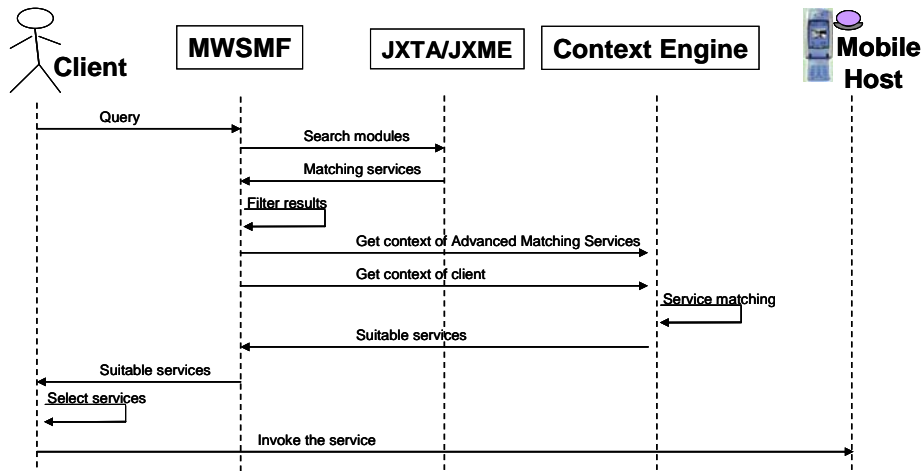

Figure 4: Mobile Web Service discovery scenario

The complete mobile Web Service discovery scenario is shown in figure 4. The client initiates the query for services at the mediation framework. The mediation framework, searches for the matching modules advertisements in the JXTA network. The module advertisements are then filtered with TF-IDF based weight of the keywords. The advanced matching services are processed at the context engine, considering the context information of the services and the client. The matching results are then forwarded back to client. The client scrolls through the list of the services and selects the best possible Web Service. The user can then access the Web Service from the Mobile Host.

## 6 Conclusion and future work

The paper addresses the concept of publishing and discovery of Web Services deployed on Mobile Hosts in peer to peer networks. Mobile Web Service provisioning and the performance analysis of the developed Mobile Host are introduced. Scope of the Mobile Host in P2P networks was later observed and the features supporting publishing and discovery the mobile Web Service in JXTA networks were mentioned. Advanced querying of the mobile Web Services with features like post filtering with weight of keywords and context-aware service discovery were addressed in detail.

The approach paves lot of scope for further research. From the technical perspective, the Mobile Host should be realized in JXTA network and the final deployment scenario is to be reached. Accessing the mobile Web Service in JXTA network, apart from the IP network is also of high interest. The approach provides alternative addressing mechanisms for Mobile Host.

The context engine and the features of the context-aware service discovery are yet to be observed. The development of Enterprise Service Bus based Mobile Web Services Mediation Framework providing the SOA features and enterprise level integration of Mobile Hosts, JXTA network, context-aware systems is also of high interest.

## Acknowledgement

The work is supported by German Research Foundation (DFG) as part of the Graduate School "Software for Mobile Communication Systems" at RWTH Aachen University.